# Upgrade or Switch: Do We Need a Next-Gen Trusted Architecture for the Internet of AI Agents?

Ramesh Raskar, Pradyumna Chari, Jared James Grogan, Mahesh Lambe, Robert Lincourt, Raghu Bala, Aditi Joshi, Abhishek Singh, Ayush Chopra, Rajesh Ranjan, Shailja Gupta, Dimitris Stripelis, Maria Gorskikh, Sichao Wang

*Project NANDA*

## Introduction

The web is on the cusp of a profound transformation. Despite advances in automation and event-driven design, the current Web still operates largely on a reactive model. Systems wait for user or client requests before acting, with limited native support for proactive or autonomous behaviors. The emerging Internet of AI Agents - a network where independently addressable software AI agents discover one another, authenticate, and act with varying degrees of autonomy - promises not only to serve human requests but to let AI agents negotiate, coordinate, and transact directly on their behalf.

Unlike traditional web components that remain idle until triggered by a user or a client issues a request, these AI agents are long-lived, goal-oriented, proactive computational entities with built-in reasoning capabilities that can anticipate needs, take initiative, maintain ongoing state, retain contextual memory and work towards defined goals without constant human direction. AI Agents leverage advanced machine learning models to interpret ambiguous instructions, adapt to changing circumstances, and make context-sensitive decisions within their domain of operation - capabilities that move far beyond the web's traditional, stateless request-response paradigm and exist on a continuum of autonomy.

AI agents, operating with varying degrees of autonomy, are poised to reshape both **human–computer interaction** and **agent-to-agent** interaction with digital systems [2] and with each other through digital intermediaries. Figure 1 contrasts today's reactive page/API model with this proactive, memory-driven architecture.

To illustrate this shift from reactive to proactive systems, consider how a human interacts with a traditional web interface versus an intelligent agent. The diagram below (Figure 1) contrasts a typical request-response interaction with a goal-driven AI agent architecture.

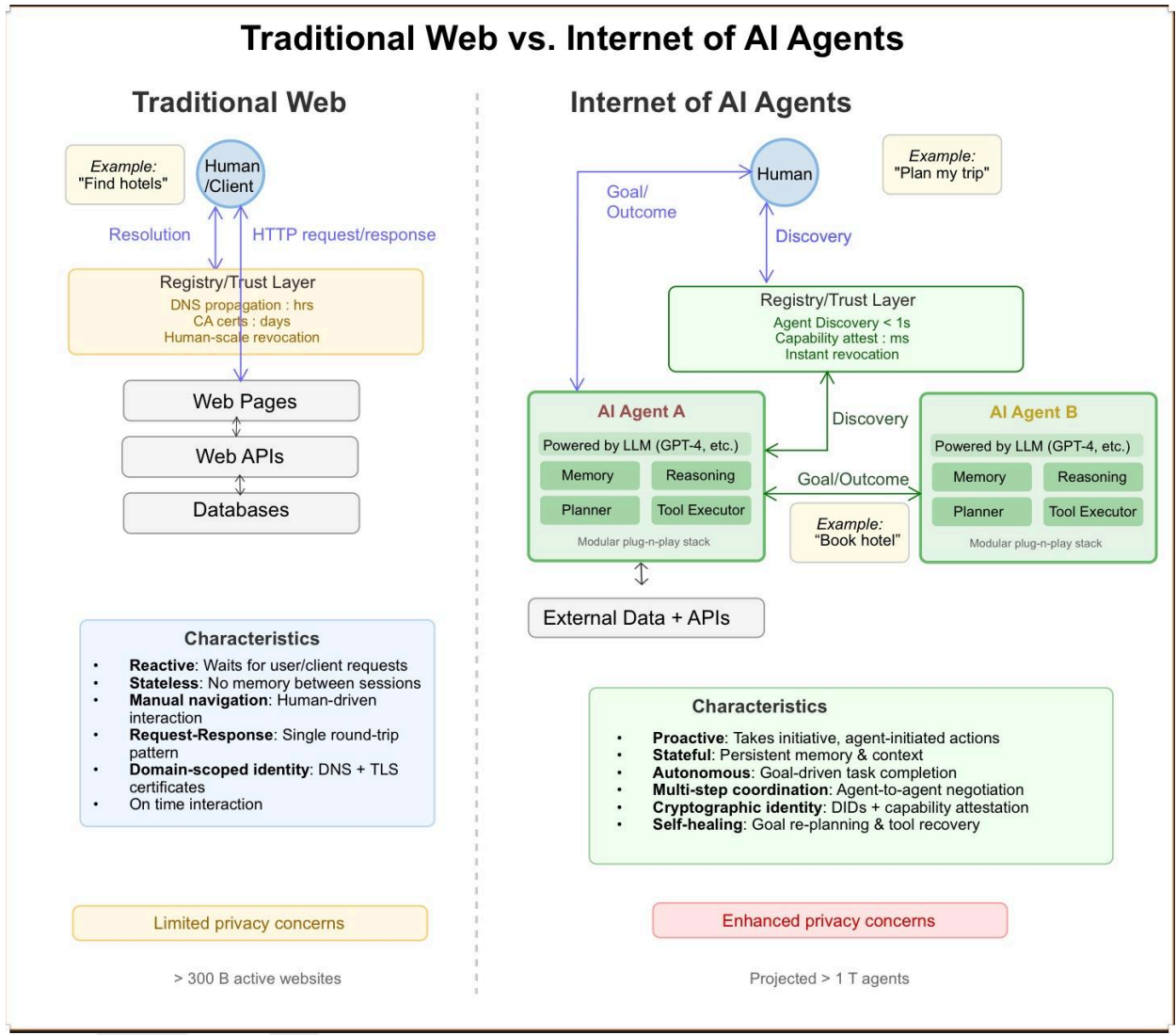

**Figure 1 | Traditional Web vs. Internet of AI Agents Architecture:** The traditional web is a **reactive request-response system**: a user (or client app) issues an HTTP call, a stateless server returns a page or JSON payload, and the interaction ends. By contrast, the Internet of AI Agents introduces **stateful, persistent, LLM-powered AI agents** that can interpret user goals, retain context across sessions, and autonomously pursue multi-step tasks,often coordinating with other agents on the user's behalf. The diagram emphasizes three structural pivots: (1) **Control-flow, origin** : human-triggered requests agent-initiated actions, (2) **Statefulness** : stateless pages/APIs → agents with durable memory embeddings, and (3) **Coordination fabric** : single request-response links → networks of cooperating agents. This migration from reactive retrieval to goal-directed execution brings new requirements for privacy, auditability, and trust, because reasoning and data aggregation increasingly occur outside an explicit human command loop.

While Figure 1 provides a high-level contrast, Table 1 formalizes this transformation by mapping the core execution properties of each system. It compares three generations of web execution,static sites, cloud APIs, and autonomous agents,across a set of key technical dimensions including statefulness, control flow, concurrency, and security posture.

To understand how this shift manifests technically, we compare three generations of web execution models,static pages, stateless APIs, and LLM-backed autonomous agents,across key dimensions like control flow, lifecycle, and trust. Table 1 summarizes this evolution:

**Table 1: Architectural comparison of execution models**

| Dimension | Static Web Page (HTML/CSS) | Stateless Cloud Function / REST API | LLM-Backed Autonomous Agent |
|---|---|---|---|
| Who initiates control? | **Client** fetches URL | **Client** calls endpoint | **Agent** decides and pushes messages; client optional |
| Lifecycle | Immutable file; versioned manually | Ephemeral (per-request); no memory between calls | Persistent process / container; maintains long-term state & memory |
| Execution context | Web server, no compute after render | Isolated runtime (e.g., AWS Lambda) spun up per call | Event loop with scheduler, tool-calling sandbox, vector store, policy layer |
| State management | External DB or none | Must externalise state every call | Internal memory + external stores; can self-modify plans |
| Autonomy level | 0 – passive | 1 – reactive | 2-3 – proactive (can set sub-goals, spawn agents) |
| Concurrency model | One request ↔ one response | Many isolated calls; no inter-call coordination | Parallel, asynchronous task graph; may coordinate with peer agents |
| Addressing & identity | DNS + TLS cert bound to domain | Same as static, plus API keys | Needs cryptographic Decentralized Identifiers (Machine Readable Identifiers) and capability attestation; identity may migrate |
| Security surface | XSS, CSRF | Injection, auth bypass | Prompt injection, tool-chain abuse, autonomous exfiltration |
| Typical latency budget | ≥ 100 ms round-trip (human perception threshold) | 10-100 ms service-to-service RPC | Internal loop: < 250 ms; External goal fulfilment: 0.25–3 s (LLM inference + network) |
| Failure semantics | 404 / 5xx | Retry logic | Must handle goal re-planning, degraded tools, dynamic trust revocation |

The table contrasts three generations of web-execution artifacts,static pages, stateless cloud functions, and LLM-backed autonomous agents,across fundamental dimensions such as who initiates control, lifecycle persistence, state handling, concurrency, identity, security posture, latency, and failure semantics. It highlights the escalating shift from passive, human-triggered resources to proactive, memory-rich agents that require new addressing, trust, and orchestration mechanismsUnlike traditional web assets,static pages or stateless API endpoints that only respond after an explicit client request,LLM-backed agents initiate control flow themselves. They persist as long-lived processes, observe their environment, and decide when and how to act. The table below isolates the key technical deltas among three execution models.

The progression from static sites to autonomous agents represents more than a technical evolution,it is a fundamental shift in agency and causality. Control no longer originates in a human request but within the agent itself. First, execution moves from user-triggered to agent-initiated. Second, memory shifts from external state to long-lived context. Third, identity becomes portable and cryptographically verified. Lastly, agents must recover from degraded tools and incomplete plans, not just retry failed calls. These changes overwhelm web infrastructure designed for passive resources,forcing a rethinking of trust, addressing, and coordination at internet scale.

Autonomous AI agents magnify the internet's scaling challenge: instead of five billion human browsers, we may soon face **trillions of always-on software actors** negotiating with one another in real-time. Originally designed for human-driven, browser-based interactions, core systems such as DNS, static IP addressing, and human-centric certificate workflows were not built to support the scale, speed, and security demands of billions of automated agents communicating and transacting in real time. As of 2025, there are over 1.1 billion websites, yet only about 193 million (17%) are actively maintained . Meanwhile, the internet serves 5.56 billion users worldwide, and there are approximately 7.21 billion smartphones in use [1] . Today, cloud providers already process **trillions of serverless invocations per month**, and IoT devices outnumber humans **three-to-one**; an agentic web will push these figures orders of magnitude higher. Without a lightweight index which encompasses a registry architecture that supports millisecond-level updates, fine-grained identity, and programmable trust, our existing stack will bottleneck innovation. This vast and rapidly growing digital ecosystem underscores the need for a more robust and scalable infrastructure to support the emerging network of autonomous agents.

This paper examines whether the path forward is: (i) a pure upgrade of the existing web infrastructure which is a DNS-centric stack, (ii) a complete switch to entirely purpose-built light weight index of registry architectures specifically designed for the 'Internet of AI agents', or (iii) a phased **hybrid** of the two. Drawing on precedents such as dialup-to-broadband and IPv4-to-IPv6 shifts, we identify the architectural layers (addressing, discovery, trust, capability attestation) where current web foundations break down and propose measurable targets  e.g., **sub-second index updates** and **cryptographically verifiable capabilities**,  for an 'Internet of AI Agents' to flourish.

**Table 2: Glossary of Key-terms**

| Term | Concise Definition | Section / Page |
|---|---|---|
| ACME (Automatic Certificate Management Environment) | IETF protocol that automates TLS-certificate issuance and renewal; modern CAs (e.g., Let's Encrypt) use it to reduce issuance to seconds-highlighting that revocation, not issuance, is the lingering bottleneck. | Upgrade Options → Certificate Management |
| Agent (Autonomous AI Agent) | Software entity with goal-directed reasoning, memory and the ability to initiate actions, migrate, or spawn helpers without continuous human supervision. | Introduction; WebPages vs Agents |
| Agent Index | Authoritative system (central, decentralised or hybrid) that stores cryptographic identifiers, capability descriptors, trust metadata and audit logs for agents. | Throughout |
| AgentFacts | Proposed metadata extension (our paper) that binds an agent's ID to capability hashes, policy constraints and runtime attestations. | Comparative Analysis |
| BGP (Border Gateway Protocol) | Internet's inter-domain routing protocol; table growth and router FIB limits are cited as a hard ceiling for per-agent IPv4/IPv6 prefixes. | Challenges in Scaling |
| Capability-Based Addressing | Discovery model that lets clients ask for a *function* (e.g., "/translate-en-es") rather than a static name; returned agents present cryptographic proof of capability. | Switch Options |
| Certificate Authority (CA) | Entity that issues X.509 certificates binding public keys to domain names/IPs; current human-oriented revocation (CRL/OCSP) is flagged as too slow for agent churn. | Primer on WWW; Upgrade Options |
| CRL / OCSP | Certificate Revocation List and Online Certificate Status Protocol-mechanisms browsers use to learn if a certificate is revoked; cannot keep pace with millisecond-level agent revocations. | Challenges in Scaling |
| DNS (Domain Name System) | Hierarchical service that maps human-readable names to IP addresses; update-propagation latency and cache churn are recurring bottlenecks. | Primer on WWW |
| DNS Push (RFC 8765) | Extension that lets resolvers subscribe to real-time updates-cited as a possible upgrade path for sub-second agent record changes. | Upgrade Options |
| DID (Decentralised Identifier) | W3C standard for self-sovereign, location-independent identifiers resolvable without a central registry; proposed in switch scenarios for agent identity. | Switch Options |
| Endpoint (Static) | Earliest stage in continuum-static web resource fetched on demand, no autonomy. | Continuum Diagram |
| IPv4 / IPv6 | 32-bit and 128-bit Internet address families; IPv4 exhaustion and IPv6 routing-table growth are discussed as address-space constraints. | Retrofit Ideas |
| Latent Capability Threshold | Point at which discovery latency, revocation speed and behavioural attestation requirements exceed the limits of legacy web protocols (between "Workers" and "Agents" in the continuum diagram). | Continuum Section |
| OCSP Stapling | TLS feature that delivers revocation proof during handshake; suggested for sub-minute validity windows in upgrade path. | Comparative Analysis |
| RDAP (Registration Data Access Protocol) | JSON-based successor to WHOIS; proposed to carry new agent capability fields and trust-score URIs. | Upgrade Options |
| Registry Shard (Private) | Enterprise-local slice of the global index/registry; resolution must query internal shard first, then cascade outward while emitting audit logs. | Search Path Configuration |
| Search Path (Configurable) | Ordered list of registries an agent must query-private → public-analogous to split-horizon DNS; necessary for policy-compliant discovery. | Search Path Configuration |
| Service | Persistent API or microservice-stage 2 in the continuum; centrally orchestrated and well-served by today's service-mesh tooling. | Continuum Section |
| SSI (Self-Sovereign Identity) | Identity model (often DID-based) where entities manage their own credentials without a central authority; appears in switch options. | Switch Options |
| SVCB/HTTPS DNS Records | Modern record types that can embed alternative endpoints and parameters; suggested for carrying agent capability hashes in an upgrade scenario. | Upgrade Options |
| Worker | Event-driven compute unit (serverless function, RPA bot)-stage 3 in continuum; identity still provisioned by DevOps pipelines. | Continuum Section |
| Zero-Knowledge Capability Proof | Cryptographic proof that an agent possesses a capability without revealing underlying data; listed among "unknown unknown" privacy techniques. | Unknown Unknowns |

## A Primer on the WWW Architecture and Hierarchy

Today's web stack hinges on four interlocking layers i.e. DNS, WHOIS, IP addressing, and Certificate Authorities, each optimised for human-initiated traffic..

**The Domain Name System (DNS)** is a globally distributed, hierarchical namespace which maps human-readable domain names to machine-readable IP addresses. This system provides globally unique identifiers for websites, a hierarchical namespace structure (root, top-level domains, second-level domains), distributed management through multiple registrars, and resolution services with propagation times typically measured in hours. While root-to-resolver updates can propagate in seconds at large providers, long-tail consumer ISPs still cache aggressively, stretching worst-case visibility to ~24 h. DNSSEC adds authenticity;but not freshness, an issue that metastasises when agents re-host every few seconds.

**The WHOIS Database** complements DNS by providing metadata about domain ownership, including contact information for domain owners, registration and expiration dates, name server information, and limited verification of identity. But after GDPR most records are redacted or proxied, and identity checks remain largely self-attested, which is an unacceptable foundation for autonomous agents that must negotiate trust without human arbitration.

**IP Addressing** provides unique identifiers for devices connected to the internet. IPv4 uses 32-bit addresses, limiting the namespace to approximately 4.3 billion addresses, while IPv6 uses 128-bit addresses, theoretically allowing for $2^{128}$ unique addresses. Yet, address *abundance* $\neq$ address *usability*. IPv6 deployment remains uneven ($\approx$ 44% of traffic) and legacy code, ACLs, and monitoring tools still assume IPv4 literals. Moreover, both IPv4 (via CGNAT) and IPv6 frequently re-assign addresses in cloud and mobile contexts, meaning an agent's IP cannot serve as a stable identifier or trust anchor.

**Certificate Authorities** issue digital certificates that authenticate website identities and enable secure communication. They validate domain ownership, issue certificates with expiration dates, maintain certificate revocation lists, and operate at human-oriented speeds and verification levels. Modern ACME workflows (e.g., Let's Encrypt) reduced certificate issuance to seconds, yet revocation and behavioural attestation remain unsolved. CRLs and OCSP struggle at today's scale; an agentic web with trillions of cert-bearing entities would need **near-instant, fine-grained revocation** that is far beyond current browser PKI.

This architecture has served the human-centric web remarkably well for decades but was designed with human timescales and interaction patterns in mind. The question now is whether this foundational architecture can evolve to support the emerging requirements of autonomous agents.

## A Lesson from Dial-up → Broadband

The transition from dialup to broadband internet provides valuable insights into how we might approach the shift to the Internet of AI agents. When the internet was first commercialized, existing telephone infrastructure seemed like a natural fit - it already connected most homes and businesses. However, as internet usage evolved, fundamental limitations of dial-up became apparent. Just as dial-up piggy-backed on copper phone lines before bandwidth demands forced DSL, cable, and eventually fiber, today's AI agents could piggy-back on DNS/IP; but AI Agents' need for millisecond-level identity, trust, discovery and coordination may ultimately demand purpose-built layers.

## Why We Didn't Use Dialup for Internet Long-Term

Dialup's limitations (max 56 kbps downstream vs. >1Mbps early DSL) revealed the importance of designing infrastructure for future needs rather than just current requirements. The circuit-switched connection model was flawed for internet use, as dial-up's temporary connections were not suitable for the persistent connectivity the internet would ultimately demand.

The addressing system also proved inadequate. Phone numbers encode geography for voice switches, but not logical topology for packet routing. Phone numbers, while hierarchically coded for voice switches, offered no stable, routable identity in an IP network, and callers could reach the ISP but not be reached in return, breaking peer-to-peer use-cases. . Additionally, the system suffered from one-way dial-out connections, where home machines could dial up the mainframes but not the other way around. Additionally, per-minute tariffs and 200ms handshake latency stifled always-on applications.

Scalability emerged as another critical limitation. The telephone system was designed for human-to-human voice communication, not for the massive data transfers that would become common with the growth of the internet. Broadband which is packet-switched, flat-rate, symmetric, and persistent, was not a luxury upgrade; it was a prerequisite for the web's rich, interactive era. The same kind of architectural ceiling now looms for AI agent discovery, identity, and trust. Broadband only exploded when new apps demanded it and AI agentic services will need a similar 'killer-app' pull.

## How We Dealt with Known Unknowns

Engineers could plainly see that dial-up's 56 kbps ceiling and ≈ 200 ms modem latency would strangle bandwidth-hungry, interactive apps. So they designed last-mile upgrades i.e. DSL, cable, then fiber and that offered > 1 Mbps downstream and sub-30 ms RTTs while staying 'always-on.'

Moving from circuit-switched voice to packet-switched data also demanded network topologies that supported many-to-many flows: early CDN experiments and P2P systems (Napster, BitTorrent) tested these limits.

Discovery evolved in parallel: search engines and the DNS hierarchy scaled from thousands of sites in 1994 to more than a billion indexed pages by 2000, proving that **addressing + search** - not telephone numbers - were the real breakthroughs. These lessons suggest that the AI agentic era will likewise require orders-of-magnitude improvements in lookup latency, persistent addressing, and capability-level discovery.

## How We Prevented Unknown Unknowns

The move to packet-switched, layered networks with TCP/IP, created a flexible foundation that could accommodate unforeseen developments in several key ways and the internet stack provided insurance against the 'unknown unknowns' of the 1990s - and paradoxically, spawned new ones. The open IPv4/IPv6 substrate allowed bandwidth-intensive applications to flourish (like YouTube, now > 60% of consumer downstream traffic) in a way that no one could have predicted, such as video and streaming. The layered protocol design enabled security features to be added without disrupting the underlying infrastructure, leading to the development of HTTPS and modern security protocols. The addressing system provided the ability for a single IP address to host multiple services and applications, enabling the growth of SaaS and diverse endpoints. Finally, the flexible addressing scheme allowed for entirely new categories of connected devices, including mobile phones ( approx. 15 Billion) and IoT devices, demonstrating how a well-designed foundation can support innovations beyond its original conception. Layering also meant we could retrofit HTTPS atop HTTP a decade later, yet that upgrade required new browsers, servers, and a global CA ecosystem, which is proof that flexibility still carries upgrade friction. Likewise, TCP/IP's openness invited spam and DDoS, illustrating that extensibility without embedded trust can backfire, underscoring the need for built-in trust primitives for AI agents.

This history demonstrates that successful infrastructure transitions require both addressing known challenges and building in flexibility for unforeseen developments. The Internet of AI agents will likely require similar foresight,particularly as billions of AI agents begin operating simultaneously across networks, creating unprecedented demands for coordination, resource allocation, and security that we cannot fully anticipate today.

The broadband transition illustrates a deeper truth: infrastructure does not just enable applications,it shapes what kinds of applications are even conceivable. Packet-switching, always-on connections, and scalable routing did not simply improve internet use; they redefined it. Similarly, the architecture we choose for AI agents registries will either constrain or unlock future capabilities and governance models for digital agents.

## Possible Extended Use of Current DNS

Before examining entirely new architectures, it's worth considering how the current DNS system might be extended to accommodate autonomous AI agents:

### Walled Gardens: Few IP Addresses but Many Endpoints

One approach would be to treat agents similarly to how web applications are currently managed: Each AI agent platform (i.e. specialized agent hosting services) would maintain a range of IP addresses. Individual AI agents would be identified by paths or subdomains (example.com/Agent1 or Agent1.example.com/). Internal routing would direct traffic to specific agents i.e. Akamai edge config pushes ~100 k rules/POP; an AI agentic platform may need 10 - 100×.

However, this approach faces several challenges. While it leverages existing infrastructure, it creates potential bottlenecks and single points of failure, as well as security and compliance issues. Each walled garden will have to agree to global protocols to talk to agents in another walled garden, which will add latency and performance variations. The walled garden hosts may throttle performance based on an opaque set of rules.

Privacy concerns also emerge with this model, because the walled-garden host retains full observability over every agent interaction. While users can readily initiate a call to *DeltaAgent1*, the reverse direction is feasible only through device-token or push-gateway intermediaries (e.g., WebPush, APNs) or other NAT-traversal techniques, adding latency and increasing platform lock-in.

While URL paths or sub-domains give a quick on-ramp, they effectively hand the hosting platform, rather than the organisation running the agent - control over naming, TLS termination, observability, and throttling. A future-proof design would assign each agent a **cryptographic identifier (e.g., DID:delta-booking-v1)** that *resolves* to its current hosting URL via a registry/index. That indirection preserves continuity when the AI agent migrates hosts, and allows trust metadata or revocation status to travel with the identifier, not the platform.

## Each Agent with an IPv4 Address

Assigning every AI agent a unique public IPv4 address (such as @Agent1 = x.y.z.w), offers superficially simple, DNS-compatible reachability, but it collapses under scrutiny. The system would be severely limited by the IPv4 address space (4.3 billion addresses, of which over 90% are already allocated according to IANA, with critical regions like APNIC and RIPE NCC having exhausted their allocations since 2011). The global free pool has been exhausted since 2011, and secondary-market /32 leases average ≈ $0.40 per month i.e. putting a trillion agents at roughly $400 billion in monthly address fees. Even if cost were ignored, the routing fabric would buckle: injecting $10^{12}$ /32 prefixes would inflate today's ~1 million-entry BGP table by six orders of magnitude, demanding tens of terabytes of high-speed TCAM in every core router. Most cloud- or mobile-hosted agents would also sit behind carrier-grade NAT, breaking inbound connectivity unless brittle hole-punch or reverse-tunnel work-arounds are used i.e. re-introducing latency and single-point failures. Security fares no better: per-agent IPv4 exposes an enormous DDoS surface, while address allocation conveys zero cryptographic proof of agent identity or code integrity. Taken together, IPv4-per-agent is not merely impractical, it is economically untenable, operationally unscalable, and architecturally incompatible with the trust requirements of an Internet of AI Agents.

## Each Agent with IPv6 Address

IPv6 offers a much larger address space, with $2^{128}$ possible addresses. This appears to solve the scarcity: one could hand every grain of sand, and every AI agent its own /128. This approach would allow directly addressable agents without intermediaries while maintaining compatibility with existing DNS systems. In practice, direct IPv6 addressing introduces four show-stoppers. **(1) Stability:** Mobile and cloud stacks use privacy extensions that rotate IPv6 interface IDs hourly to daily; a literal address is therefore not a durable identifier. **(2) Routing scale:** Today's global BGP table holds ~1 M IPv6 prefixes. Advertising even $10^9$ agent-specific /128s would $10^6\times$ inflate router FIB/RIB memory, far beyond hardware limits. **(3) Registry/Index bloat:** Reverse-DNS and DNSSEC signatures for billions of PTR records would add petabytes of zone data and hours of SOA transfer time, defeating sub-second lookup goals. **(4) Trust signal void:** An IPv6 literal conveys no ownership or code-integrity proof, forcing an extra identity layer anyway. Additionally, this requires full IPv6 adoption, which has been slow over time (global IPv6 adoption stands at approximately 44%, with significant variation across countries [5]); We have an address space that is abundant but operationally unstable, economically burdensome, and trust-agnostic - useful as a transport plumbing, but insufficient as a standalone agent identifier.

## An Important Consideration: Boundary-Aware Naming and Configurable Search Paths

An essential capability that must be addressed is the configurable resolution paths for AI agents, analogous to split-horizon DNS. Traditional DNS allows seamless movement between intranet and internet resources, automatically routing queries through organizational internal name servers and only on a miss, forwards the query to public roots for resolution - seamlessly bridging intranet and internet. However, no comparable mechanism exists for AI agent discovery: - organizations cannot effectively intersect their internal search index with public search indexes without leaking data or losing policy control.

For the Internet of AI agents to succeed behind the firewall, each agent must honor an administrator-defined resolution order that prioritizes internal registries, then selectively consults external registries. . Organizations need the ability to prioritize internal agent discovery and communication while maintaining controlled access to public domain agents. At the same time, every cross-boundary lookup must emit a verifiable audit event, so security teams can trace when an AI agent leaves the private domain and verify that it adhered to data-handling policies. Without policy-aware resolution and boundary-crossing traceability, enterprises will not risk deploying agents that must collaborate both inside and outside their networks.

## From Static Endpoints to Services to Workers to Autonomous Agents: a Continuum of Web Actors

When the Web is described as moving "from web pages to agents," it is easy to imagine a sudden, binary leap. In practice the evolution is gradual, and most of the infrastructure we rely on today was stretched, sometimes painfully, at each intermediate step. By recognising this continuum we can isolate the exact point at which the legacy stack finally breaks, and therefore justify why a new agent-oriented index is required.

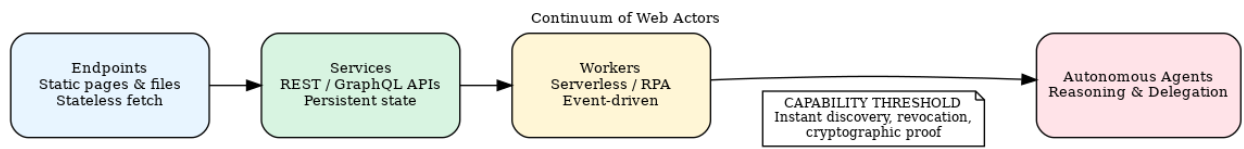

### 1. Endpoints : The Stateless Web

The original Web exposed **static endpoints** i.e. HTML files, images, style sheets and, later, simple CGI scripts. A user (or crawler) made an explicit request, the server returned a byte stream, and the conversation ended. DNS, HTTP caching hierarchies and domain-validated TLS certificates were more than adequate for these one-shot interactions; nothing in the model required rapid renaming, mid-flight revocation, or complex policy checks.

**2. Services : Always-On APIs**

The next layer of capability arrived when businesses wrapped their databases in **REST and GraphQL services**. Now machines, not just humans, were first-class clients, and data mutated continuously rather than on release nights. We coped by bolting on *service discovery* (SRV records, load balancers, Envoy side-cars) and by hardening transport with mutual-TLS. Yet the architecture remained centrally orchestrated: DevOps pipelines created each service, security teams issued its credentials, and version upgrades rolled out on a schedule measured in hours or days.

**3. Workers : Event-Driven Compute**

Cloud platforms then introduced **serverless functions, cron jobs and RPA bots**. These "workers" are short-lived but bursty; thousands can spin up when a message queue spikes and disappears a second later. The industry responded with function gateways, granular IAM roles and CI/CD automation. Even so, the identity of every worker is still provisioned *by humans* (or human-authored pipelines), and revocation can tolerate minute-scale propagation. The web's naming, security and governance fabric is stretched, but it holds.

**4. Agents : Autonomy and Delegation**

**Autonomous agents** sit at the far end of this continuum. They do not merely react to events; they pursue goals, maintain memories, migrate between runtimes and, crucially, **delegate sub-tasks to freshly spawned helper agents**. Their population can rise or fall by millions in the time it takes a Kubernetes deployment to roll one replica forward. At this stage three new thresholds appear:

1. **Self-directed discovery**
   An agent must discover, evaluate and negotiate with *unknown* peers in **milliseconds**.
   Static DNS records, manual API catalogues and vendor-specific service meshes cannot supply such real-time, cross-domain capability lookup.

2. **Delegated authority with rapid revocation**
   When an agent hands a helper a subset of its privileges, the grant must be revocable *instantly* if the helper misbehaves. OAuth tokens that linger for minutes and X.509 certificates that cache for hours are no longer acceptable.

3. **Cryptographic proof of behaviour**
   Trust can no longer hinge on "I control this domain." Regulators and counterpart agents will demand **code-integrity attestations** and **tamper-evident execution logs** that travel with the agent wherever it migrates.

These thresholds are qualitative, not incremental: they render the legacy stack's assumptions i.e. slow propagation, human-issued credentials, domain-scoped identity that is fundamentally obsolete.

To understand why current architectures may be insufficient, we must recognize fundamental differences between static web resources and autonomous AI agents:

**Traditional Web pages** are primarily passive resources where content is served upon request, interactions are limited and predefined, changes are typically made by human administrators, and identity is relatively stable and long-lived. Even modern push features (WebSockets, SSE, serverless functions) remain centrally orchestrated and human-triggered.

**AI Agents**, by contrast, are dynamic, evolving programs. They initiate actions and requests and can be ephemeral, appearing and disappearing in the orders of magnitude more frequently than website deployments.. They make autonomous decisions, may change behavior over time, and may create other agents. Additionally, they may move between hosting environments, many will consume non-trivial compute, ranging from small WASM sandboxes to GPU-bound planners.

## Potential Issues

These differences create several potential issues when trying to use current web infrastructure:

1. **Latency** - Agents demand millisecond-level capability look-ups, key exchange, and revocation; human-tolerant DNS or OCSP latencies are orders of magnitude too slow.
2. **Security & Privacy** - New vectors: prompt-injection, sybil swarms, plug-in supply-chain poisoning; mandate signed policies and runtime sandboxing.
3. **Traceability** - The autonomous and potentially self-modifying nature of AI agents complicates tracking their decision-making processes, making accountability and compliance more challenging. It requires append-only, tamper-evident execution logs to prove compliance.
4. **Verification** - Beyond domain-validated TLS, AI agents need cryptographic code-integrity attestations and verifiable-revocable capability tokens, refreshed in seconds, not days.
5. **Scalability & Coordination:** The Internet of AI Agents enables distributed coordination and adaptive task routing, offering resilience and scalability. Billions of short-lived actors need decentralised, locality-aware routing, not central schedulers.
6. **Data & Application Models:** The classic web relies on rigid REST APIs and siloed data with limited interoperability, while agentic systems embrace semantic web standards, support dynamic interactions, and utilize machine-readable intents for seamless negotiation between AI agents.
7. **Governance & Quota:** Autonomous compute must respect resource budgets and kill-switch policies enforced at the registry/index layer.

These fundamental differences suggest that simply extending current systems may not be sufficient for the needs of the Internet of AI agents. New aspects, such as the dynamic verification of behavior, traceable decision histories, and revocable privileges must be paired with behavioral metadata, reputational scores, or cryptographic attestations to allow systems and users to assess agent reliability in real-time.

## Challenges in Scaling - Unique Crossover Points

As we consider the transition to an Internet of AI Agents, three categories of scaling challenges emerge i.e. addressing & routing, real-time trust propagation, and governance; each a critical 'crossover point' where legacy web systems begin to fail.

### Known Unknowns

Several critical technical limitations are already visible when considering the current web infrastructure for the Internet of AI Agents. Address scarcity remains a constraint despite NAT and IPv6 (i.e. IPv6 offers $2^{128}$ space); but the bottleneck is routing-table inflation and privacy-rotating IPv6 addresses churn, not mere numeric supply Additionally, IPv6 adoption remains uneven and slow adoption rate over time (~ 44 % of global traffic as of May 2025).

**DNS update propagation** poses another significant challenge [3]. End‑user visibility is gated by **public recursive resolvers**: some consumer ISPs cache records aggressively or ignore low‑TTL settings, so worst‑case propagation across the full resolver ecosystem can still stretch to **24 -48 hours** [4], which becomes unworkable for the Internet of AI agents with dynamic capabilities and locations. Agent index must converge while handling billions of writes per hour.

**Trust metadata** is another gap. RDAP replaces WHOIS with JSON that is minimal and oriented toward human ownership, while AI agents will require rich metadata about capabilities, permissions, ownership, code-integrity digests, and revocation status and other characteristics. Existing systems have no standard way to represent or query this information.

**Certificate lifecycle** reveals the additional bottleneck. ACME can issue a cert in under a minute, yet CRL/OCSP mechanisms cannot yank or refresh trillions of certificates in real time, and they convey nothing about an AI agent's behavioural guarantees like creation, modification, and verification. Additionally, certificate revocation lists would become unmanageably large [8].

### Unknown Unknowns

**Unknown Unknowns** still loom over an Internet of AI Agents, even after we solve for address space, metadata, and issuance speed. **Latency** requirements and **security** models raise fundamental questions about how quickly (i.e. 5 ms, 50 ms or 500 ms) complex AI agents will need to discover globally and authenticate other agents using cryptographic attestation to complete and remain stable. Every extra round-trip or zero-knowledge proof increases safety but consumes the latency budget; empirical thresholds have yet to be measured and what new attack vectors will emerge.

**Governance** and **interoperability at a trillion scale,** present equally complex challenges. ICANN's single-root model and W3C's DID registries provide templates, but no consensus exists on how competing registries or sovereign states will federate trust for autonomous software. We must determine how standards will be established and enforced across potentially trillions of agents, and how agents from different frameworks and with different capabilities will communicate effectively. A polycentric model may emerge, or the ecosystem may fragment into incompatible "agent nets."

**Privacy** under continuous delegation is particularly significant. The protocols and standards that will ensure robust data minimization and privacy in the Internet of AI agents remain undefined, raising open questions about how user data can be protected across billions of autonomous AI agents. Techniques such as zero-knowledge capability proofs, differential privacy, or unlinkable yet accountable identifiers are promising but untested at planetary scale.

**Compute Confidentiality & Data Locality** present emerging needs for privacy-preserving compute and federated data access, raising questions about the adoption of secure enclaves, homomorphic encryption, and other advanced storage models; yet remain constrained by hardware limits i.e. full homomorphic encryption is orders-of-magnitude slower than plaintext compute. Request handlers face a related challenge: handling billions of real-time agentic requests will require breakthroughs in scalable, distributed load balancing and intent-aware orchestration. Whether hybrid models like stateless ZK-rollups for proofs plus enclave execution for secrets, can meet both performance and compliance targets is unclear. .

**Intent-Aware Orchestration** is another challenge i.e. routing large scale real-time AI agent requests cannot rely on today's L4/L7 load balancers. Schedulers must become intent-aware, matching tasks to agents based on capability, jurisdiction, and carbon cost, yet no production-grade algorithms or economic models exist. We may discover that entirely new coordination primitives are required, or that incremental extensions of service-mesh technology suffice; at present, the answer is unknowable.

## Upgrade or Switch

Given these challenges, we face a fundamental decision: should we upgrade existing web stack or switch to purpose-built registries designed specifically for the Internet of AI agents, or adopt a hybrid approach that blends both?

### Upgrade Options

Upgrade options build upon the existing web stack rather than replacing it.

**IPv6 Dual-Stack with Agent-Aware Routing:** First, **switching to IPv6** would leverage its massive address space while maintaining compatibility with existing DNS infrastructure, providing a gradual transition path from current systems. However, per-agent /128 announcements would inflate today's BGP table by six orders of magnitude, and rotating privacy IIDs undercut identity stability. Any upgrade must therefore pair IPv6 with **aggregated, signed routing manifests** or overlay rendezvous services that shield the DFZ from per-agent noise.

**RDAP Metadata Extension:** Updating RDAP/WHOIS metadata represents another evolutionary approach. This would involve extending existing WHOIS databases with agent-specific fields, adding capability descriptions, trust metrics, and other agent-specific data, while standardizing query methods for agent-specific attributes. Please note that the ownership-only model must be expanded to include capability descriptors and code-integrity digests.

**ACME-Plus Certificates with Instant Revocation:** Automated certificate methods automates certificate issuance, but OCSP/CRL revocation still propagate in minutes; agent registries need millisecond revocation and runtime-behaviour attestations. This would include creating agent-specific trust metrics and verification standards, as well as implementing near real-time certificate revocation mechanisms. Upgrading PKI therefore means binding each certificate to a signed Software Bill of Materials (SBOM) and replacing CRL/OCSP with short-lived, stapled proofs that expire in milliseconds.

**Agent-Specific DNS Records and Push Updates:** DNS can be stretched by introducing SVCB/HTTPS records whose *svcparam* fields carry capability hashes and policy pointers. Coupled with DNS Push (RFC 8765) or DoH subscription streams, this enables near-instant propagation while remaining backward-compatible with existing resolvers, provided caches respect sub-second TTLs and deploy cache-poison safeguards. DNS automation offers a final path forward, dramatically improving DNS propagation times, implementing agent-specific DNS record types, and creating specialized agent directory services within the existing DNS hierarchy.

## Switch Options

**Clean-Slate Cryptographic Namespace:** More revolutionary approaches could fundamentally reimagine agent addressing. Creating a **new addressing system** for agents would involve developing a parallel addressing system specifically for agents, designed for agent-specific requirements from the ground up, potentially implementing entirely new resolution protocols which resolves identifiers of every AI agent through a purpose-built, millisecond-latency lookup service rather than DNS.

**Self-Sovereign DID Mesh:** Decentralized identity systems offer another paradigm shift, implementing DID (Decentralized Identifier) based systems that allow agents to maintain self-sovereign identity and enable direct agent-to-agent verification without touching a central index/registry, yet the ecosystem's 150+ 'DID methods' risk fragmentation and slow discovery; a **resolution super-router** may still be needed. [6].

**Tiered Hybrid index:** Hybrid index systems could balance centralized and decentralized approaches by maintaining centralized registries for safety-critical or high-trust agents, while using a federated mesh (DID, IPFS, or ActivityPub) for the long tail of specialized AI agents, creating bridge gateway protocols between different index systems - which sign and cache proofs, so agents can traverse tiers in <50 ms.

**Capability-First Addressing:** Capability-based addressing represents perhaps the most radical departure from current systems. This approach would address agents by their capabilities rather than static identifiers, enable dynamic discovery based on required functions, and implement semantic addressing systems. For example, a travel planner would query for '/translate-en-es' or '/optimize-route' and receive a ranked list of agents that cryptographically attest to those capabilities. Such a semantic index must include **proof-of-capability tokens** to prevent spam or spoofing.

**Table 3: Comparative Analysis**

| Design Dimension | Enhanced-Upgrade Path | Clean-Switch Path | Key Trade-off |
|---|---|---|---|
| **Identifier Space & Routing** | Dual-stack IPv6 + aggregated signed routing manifests to shield DFZ from /128 noise | Hash-derived, location-independent IDs resolved by a new overlay (e.g., Kademlia-style DHT) | Upgrade preserves BGP tooling but risks FIB bloat; switch avoids BGP but needs new resolver adoption |
| **Update / Revocation Latency** | DNS Push + DoH streams; target < 1 s global convergence; OCSP stapling with 1-minute TTL | Gossip-based or CRDT ledger with millisecond write propagation and automatic tombstoning | Upgrade easier to deploy; switch offers lower worst-case latency if overlay is well-peered |
| **Identity & Trust** | ACME-plus certs bound to SBOM digests; RDAP fields for capability & trust-score URIs | Self-sovereign DIDs + verifiable credentials; transport still uses TLS (or QUIC) | Upgrade piggy-backs on browsers; switch removes central CA reliance but fragments trust anchors |
| **Capability Discovery** | DNS SVCB/HTTPS records hold signed capability hashes; optional SRV fall-back | Capability-first queries ("/translate-en-es") resolved by semantic index with ZK-proof of capability | DNS option benefits from ubiquity but limited expressiveness; capability-index richer but new protocol |
| **Governance Model** | Extend ICANN/SSAC + IETF drafts | Polycentric federation of registries with on-chain transparency logs | Upgrade reuses existing policy channels; switch encourages innovation but risks fragmentation |
| **Implementation Timeline** | 12-18 months for global IPv6 + DNS Push rollout; incremental RDAP fields | 3-5 years to spec, standardise and bootstrap overlay; early adopters in closed ecosystems | Time-to-value vs. architectural purity |

| **Backward Compatibility** | High: HTTP(S) stacks, browsers, firewalls remain unchanged | Medium: gateways can bridge, but native support needed for full feature set | Compatibility vs. feature richness |
|---|---|---|---|
| **Operations & Cost** | Leverages existing DNS/BGP tooling; OPEX rises with churn mitigation | New infra but simpler core (no multi-gig FIB); CAPEX high at bootstrap | OPEX vs. CAPEX balance |

## Conclusion

The transition to the Internet of AI agents represents a fundamental shift in how we interact with digital systems, comparable to the shift from dialup to broadband internet. While upgrading existing systems offers a path of least resistance and backwards compatibility, the unique requirements of autonomous agents suggest that entirely new index architectures may ultimately be necessary.

The history of technology transitions suggests that hybrid approaches often emerge during periods of rapid change. We may see centralized registries for critical agents alongside decentralized systems for specialized agents, with bridge protocols enabling interoperability.

What's clear is that this infrastructure question must be addressed proactively rather than reactively. The decisions made today about agent index architecture will shape the capabilities, security, and accessibility of the Internet of AI agents for decades to come. By learning from past transitions and anticipating future needs, we can build infrastructure that not only addresses current requirements but remains flexible enough to accommodate the unknown unknowns that will inevitably emerge as autonomous agents become an integral part of our digital landscape.

Rather than simply extending human-oriented web infrastructure, we have an opportunity to design systems specifically for agent-to-agent interactions, potentially unlocking entirely new categories of applications and services. Whether through upgrade or switch - or most likely, some combination of both - the index architecture for the Internet of AI agents will be a critical foundation for the next era of digital innovation. By acknowledging that infrastructure defines both capabilities and accountability, we can create an Internet of AI agents that is not only scalable and performant, but also responsible and resilient.